\documentstyle[11pt]{article}

\renewcommand{\baselinestretch}{1.2}
\newcommand{\be}{\begin{equation}}
\newcommand{\ee}{\end{equation}}
\newcommand{\bea}{\begin{eqnarray}}
\newcommand{\eea}{\end{eqnarray}}
\newcommand{\gi}{gauge-invariant }
\newcommand{\gic}{gauge-invariant, }
\newcommand{\gis}{gauge-invariant. }
\newcommand{\N}{{\cal N}}
\newcommand{\tfrac}[2]{\textstyle{\frac{#1}{#2}}}

\begin{document}

\begin{titlepage}

\vspace*{2cm}
\begin{center}

{\large\bf
GAUGE-INVARIANT VARIABLES AND MANDELSTAM CONSTRAINTS
IN SU(2) GAUGE THEORY
\\}
\vspace*{1cm}

{\bf N. J. Watson} \\
Centre de Physique Th\'eorique, CNRS - Luminy, \\
F-13288 Marseille Cedex 9, France. \\
email: watson@cptsu1.univ-mrs.fr \\

\vspace*{1.5cm}
{\bf   Abstract  \\ }

\end{center}

The recent solution of the Mandelstam constraints for SU(2)
is reviewed. This enables the subspace of physical configurations
of an SU(2) pure gauge theory on the lattice (introduced
solely to regulate the number of fields) with $3\N$ physical degrees
of freedom to be fully described in terms of $3\N$ gauge-invariant
continuous loop variables and $\N -1$ gauge-invariant discrete
$\pm 1$ variables. The conceptual simplicity of the solution
and the essential role of the discrete variables are emphasized.

\vspace*{1cm}
\begin{center}
Talk presented at QCD '94, Montpellier, France, 7-13 July 1994.
\end{center}

\vspace*{1cm}

\noindent
August 1994

\noindent
CPT-94/P.3065

\vfill

\end{titlepage}

{\bf 1. Introduction}

\noindent
When attempting to understand the dynamics of some particular physical
system, very often the first step is choose a set of variables
appropriate to the symmetries of the problem. For a system described
by a gauge theory, the fundamental symmetry is the system's invariance
under local gauge transformations. It is therefore to be expected
that such
a system is most naturally described by a set of variables which
themselves are gauge-invariant. In particular, such a set of variables
would be expected to characterize best the intrinsically non-linear
nature of a non-abelian theory, and perhaps also,
through the resulting form taken by the Hamiltonian, to point the way
to new
non-perturbative analytic and semi-analytic calculational methods.
Given the importance of trying to understand the non-perturbative
dynamics of non-abelian gauge theories, these two reasons mean that
the study of such variables is potentially very profitable.
Here it is described how, by solving fully
the SU(2) Mandelstam constraints, such a complete
set of \gi variables can be constructed for a pure SU(2)
lattice gauge theory \cite{watson}.
The reason for working on the lattice here is that, by
regulating the  number of variables, it allows very explicict
control over them.

{\bf 2. Gauge-invariant variables?}

\noindent
In a lattice formulation of SU(2) gauge theory, the number
$3\N$ of physical degrees of freedom, obtained by dividing out the
gauge freedom of the system from the overall freedom,
is given by the dimension of the quotient space
$\otimes_{{\rm links}}{\rm SU}(2)/\otimes_{{\rm sites}}{\rm SU}(2)$.
Thus, for
a $d$-dimensional lattice with $n^{d}$ sites and periodic boundary
conditions,
\be
3\N
=
3(d-1)n^{d}
\ee

A common procedure in formulations involving loop variables is to
construct a set of basis matrices which are ``as \gi as
possible'' and from which any loop
can be formed by taking the trace of the appropriate product of
these matrices and their inverses. For a lattice of
any dimension $d$, the number of required basis matrices is $\N +1$.
An example of such a basis set for a $3\times 3$ lattice (for which
$\N + 1 = 10$) is shown in fig. 1. In this example,
${\rm Tr}\,U_{1}$ and ${\rm Tr}\,U_{2}$ are Polyakov loops
which distinguish configurations differing only by global
transformations depending on the $Z_{2}$ centre of SU(2).

These basis matrices are invariant under local gauge
transformations everywhere except at the
site from which the ``tails'' originate. Though such sets
of basis matrices have
been used as variables, e.g. \cite{mullerruhl},
they suffer from three intimately related
drawbacks: i) they are not fully \gi ii) they involve $3\N + 3$
continuous variables rather than the correct number $3\N$ iii) they
are highly non-local. The question arises, is it possible to
find a complete set of $3\N$ independent Wilson and Polyakov loop
(i.e. trace) variables
in terms of which all other such loops may be expressed?

\noindent
\begin{picture}(425,270)(0,10)

\put(5,200){\line(0,1){60}}
\put(25,200){\line(0,1){60}}
\put(45,200){\line(0,1){60}}
\put(5,200){\line(1,0){60}}
\put(5,220){\line(1,0){60}}
\put(5,240){\line(1,0){60}}
\multiput(5,200)(0.5,0){121}{\circle*{1.8}}
\multiput(40,200)(-0.5,0.5){7}{\circle*{1}}
\multiput(40,200)(-0.5,-0.5){7}{\circle*{1}}

\put(30,180){\makebox(0,0)[l]{$U_{1}$}}

\put(95,200){\line(0,1){60}}
\put(115,200){\line(0,1){60}}
\put(135,200){\line(0,1){60}}
\put(95,200){\line(1,0){60}}
\put(95,220){\line(1,0){60}}
\put(95,240){\line(1,0){60}}
\multiput(95,200)(0,0.5){121}{\circle*{1.8}}
\multiput(95,235)(-0.5,-0.5){7}{\circle*{1}}
\multiput(95,235)(0.5,-0.5){7}{\circle*{1}}

\put(120,180){\makebox(0,0)[l]{$U_{2}$}}

\put(185,200){\line(0,1){60}}
\put(205,200){\line(0,1){60}}
\put(225,200){\line(0,1){60}}
\put(185,200){\line(1,0){60}}
\put(185,220){\line(1,0){60}}
\put(185,240){\line(1,0){60}}
\multiput(185,200)(0.5,0){41}{\circle*{1.8}}
\multiput(205,200)(0,0.5){41}{\circle*{1.8}}
\multiput(185,220)(0.5,0){41}{\circle*{1.8}}
\multiput(185,203)(0,0.5){35}{\circle*{1.8}}
\multiput(200,200)(-0.5,0.5){7}{\circle*{1}}
\multiput(200,200)(-0.5,-0.5){7}{\circle*{1}}

\put(210,180){\makebox(0,0)[l]{$U_{3}$}}

\put(275,200){\line(0,1){60}}
\put(295,200){\line(0,1){60}}
\put(315,200){\line(0,1){60}}
\put(275,200){\line(1,0){60}}
\put(275,220){\line(1,0){60}}
\put(275,240){\line(1,0){60}}
\multiput(275,200)(0.5,0){81}{\circle*{1.8}}
\multiput(315,200)(0,0.5){41}{\circle*{1.8}}
\multiput(295,220)(0.5,0){41}{\circle*{1.8}}
\multiput(295,203)(0,0.5){35}{\circle*{1.8}}
\multiput(275,203)(0.5,0){41}{\circle*{1.8}}
\multiput(310,200)(-0.5,0.5){7}{\circle*{1}}
\multiput(310,200)(-0.5,-0.5){7}{\circle*{1}}

\put(300,180){\makebox(0,0)[l]{$U_{4}$}}

\put(365,200){\line(0,1){60}}
\put(385,200){\line(0,1){60}}
\put(405,200){\line(0,1){60}}
\put(365,200){\line(1,0){60}}
\put(365,220){\line(1,0){60}}
\put(365,240){\line(1,0){60}}
\multiput(365,200)(0.5,0){121}{\circle*{1.8}}
\multiput(425,200)(0,0.5){41}{\circle*{1.8}}
\multiput(405,220)(0.5,0){41}{\circle*{1.8}}
\multiput(405,203)(0,0.5){35}{\circle*{1.8}}
\multiput(365,203)(0.5,0){81}{\circle*{1.8}}
\multiput(420,200)(-0.5,0.5){7}{\circle*{1}}
\multiput(420,200)(-0.5,-0.5){7}{\circle*{1}}

\put(390,180){\makebox(0,0)[l]{$U_{5}$}}

\put(5,80){\line(0,1){60}}
\put(25,80){\line(0,1){60}}
\put(45,80){\line(0,1){60}}
\put(5,80){\line(1,0){60}}
\put(5,100){\line(1,0){60}}
\put(5,120){\line(1,0){60}}
\multiput(5,80)(0,0.5){81}{\circle*{1.8}}
\multiput(5,120)(0.5,0){41}{\circle*{1.8}}
\multiput(25,100)(0,0.5){41}{\circle*{1.8}}
\multiput(8,100)(0.5,0){35}{\circle*{1.8}}
\multiput(8,80)(0,0.5){41}{\circle*{1.8}}
\multiput(20,100)(-0.5,0.5){7}{\circle*{1}}
\multiput(20,100)(-0.5,-0.5){7}{\circle*{1}}

\put(30,60){\makebox(0,0)[l]{$U_{6}$}}

\put(95,80){\line(0,1){60}}
\put(115,80){\line(0,1){60}}
\put(135,80){\line(0,1){60}}
\put(95,80){\line(1,0){60}}
\put(95,100){\line(1,0){60}}
\put(95,120){\line(1,0){60}}
\multiput(95,83)(0.5,0){41}{\circle*{1.8}}
\multiput(115,83)(0,0.5){75}{\circle*{1.8}}
\multiput(115,120)(0.5,0){41}{\circle*{1.8}}
\multiput(135,100)(0,0.5){41}{\circle*{1.8}}
\multiput(118,100)(0.5,0){35}{\circle*{1.8}}
\multiput(118,80)(0,0.5){41}{\circle*{1.8}}
\multiput(95,80)(0.5,0){47}{\circle*{1.8}}
\multiput(130,100)(-0.5,0.5){7}{\circle*{1}}
\multiput(130,100)(-0.5,-0.5){7}{\circle*{1}}

\put(120,60){\makebox(0,0)[l]{$U_{7}$}}

\put(185,80){\line(0,1){60}}
\put(205,80){\line(0,1){60}}
\put(225,80){\line(0,1){60}}
\put(185,80){\line(1,0){60}}
\put(185,100){\line(1,0){60}}
\put(185,120){\line(1,0){60}}
\multiput(185,83)(0.5,0){81}{\circle*{1.8}}
\multiput(225,83)(0,0.5){75}{\circle*{1.8}}
\multiput(225,120)(0.5,0){41}{\circle*{1.8}}
\multiput(245,100)(0,0.5){41}{\circle*{1.8}}
\multiput(228,100)(0.5,0){35}{\circle*{1.8}}
\multiput(228,80)(0,0.5){41}{\circle*{1.8}}
\multiput(185,80)(0.5,0){87}{\circle*{1.8}}
\multiput(240,100)(-0.5,0.5){7}{\circle*{1}}
\multiput(240,100)(-0.5,-0.5){7}{\circle*{1}}

\put(210,60){\makebox(0,0)[l]{$U_{8}$}}

\put(275,80){\line(0,1){60}}
\put(295,80){\line(0,1){60}}
\put(315,80){\line(0,1){60}}
\put(275,80){\line(1,0){60}}
\put(275,100){\line(1,0){60}}
\put(275,120){\line(1,0){60}}
\multiput(275,80)(0,0.5){121}{\circle*{1.8}}
\multiput(275,140)(0.5,0){41}{\circle*{1.8}}
\multiput(295,120)(0,0.5){41}{\circle*{1.8}}
\multiput(278,120)(0.5,0){35}{\circle*{1.8}}
\multiput(278,80)(0,0.5){81}{\circle*{1.8}}
\multiput(290,120)(-0.5,0.5){7}{\circle*{1}}
\multiput(290,120)(-0.5,-0.5){7}{\circle*{1}}

\put(300,60){\makebox(0,0)[l]{$U_{9}$}}

\put(365,80){\line(0,1){60}}
\put(385,80){\line(0,1){60}}
\put(405,80){\line(0,1){60}}
\put(365,80){\line(1,0){60}}
\put(365,100){\line(1,0){60}}
\put(365,120){\line(1,0){60}}
\multiput(365,83)(0.5,0){41}{\circle*{1.8}}
\multiput(385,83)(0,0.5){115}{\circle*{1.8}}
\multiput(385,140)(0.5,0){41}{\circle*{1.8}}
\multiput(405,120)(0,0.5){41}{\circle*{1.8}}
\multiput(388,120)(0.5,0){35}{\circle*{1.8}}
\multiput(388,80)(0,0.5){81}{\circle*{1.8}}
\multiput(365,80)(0.5,0){47}{\circle*{1.8}}
\multiput(400,120)(-0.5,0.5){7}{\circle*{1}}
\multiput(400,120)(-0.5,-0.5){7}{\circle*{1}}

\put(390,60){\makebox(0,0)[l]{$U_{10}$}}

\put(0,20){\makebox(0,0)[l]{\footnotesize{
Fig. 1. A possible set of 10 basis matrices for a $3\!\times\! 3$
lattice with periodic boundary conditions.
}}}

\end{picture}

{\bf 3. Solving the SU(2) Mandelstam constraints}

\noindent
For any given SU($N$) gauge group,
even on a finite lattice the number of distinct loops which
can be formed is infinite. This implies the existence of
dependences among the loops. These are
usually known as Mandelstam constraints
\cite{mandelstam1}\cite{mandelstam2}, and are
complicated non-linear identities
stemming just from the basic properties of SU($N$) matrices.
For SU(2), which is the simplest case, they may all be derived
\cite{loll} from the fundamental identity
for arbitrary SU(2) matrices $U_{A}$, $U_{B}$
\be\label{3.1}
{\rm Tr}\,U_{A}U_{B}
-{\rm Tr}\,U_{A}\,{\rm Tr}\,U_{B}
+ {\rm Tr}\,U_{A}U_{B}^{-1}
=
0
\ee
A simple example of a Mandelstam constraint for SU(2) Wilson loops,
in this case coming directly
from eq. (\ref{3.1}), is shown in fig. 2.

\noindent
\begin{picture}(425,130)(10,10)

\put(60,69){\line(0,1){21}}
\put(60,69){\line(1,0){39}}
\put(60,90){\line(1,0){41}}
\put(99,69){\line(0,-1){19}}
\put(101,71){\line(0,1){19}}
\put(101,71){\line(1,0){19}}
\put(99,50){\line(1,0){21}}
\put(120,50){\line(0,1){21}}

\put(145,70){\makebox(0,0){${\displaystyle - }$}}

\put(170,71){\framebox(39,19){}}
\put(211,50){\framebox(19,19){}}

\put(255,70){\makebox(0,0){${\displaystyle + }$}}

\put(280,70){\line(1,0){60}}
\put(280,70){\line(0,1){20}}
\put(280,90){\line(1,0){40}}
\put(320,50){\line(0,1){40}}
\put(320,50){\line(1,0){20}}
\put(340,50){\line(0,1){20}}

\put(363,70){\makebox(0,0)[l]{${\displaystyle = \hspace{15pt}0 }$}}

\put(10,10){\makebox(0,0)[l]{\footnotesize{
Fig. 2. A simple example of a Mandelstam constraint for Wilson
loops in SU(2) lattice gauge theory.
}}}

\end{picture}

For any given set of SU(2) matrices, the
set of Mandelstam constraints among the
traces can be resolved into six distinct types of identity \cite{loll}.
In order to construct a complete set of
independent trace variables from this
infinite set of traces, it is necessary to solve
completely these constraints. At first sight, the non-linear
identities appear so complicated that to solve them
seems a hopeless task.
However, this is not the case, and their solution is in fact
conceptually very simple \cite{watson}.

The key to solving the Mandelstam constraints for a given SU(2)
lattice gauge theory is to take the previously considered
basis matrices $U_{\alpha}$, $\alpha = 1,2\ldots\N+1$,
and write them in the form
\be
U_{\alpha}
=
\alpha_{0}I + i\mbox{\boldmath $\alpha\cdot\sigma$}
\ee
where
$\mbox{\boldmath $\sigma$} = (\sigma_{1},\sigma_{2},\sigma_{3})$ are the
three Pauli matrices and the real numbers
$\alpha_{i}$, $i = 0,1,2,3$, satisfy
$\alpha_{0}^{2} + \mbox{\boldmath $\alpha\cdot\alpha$} = 1$
where
$\mbox{\boldmath $\alpha$} = (\alpha_{1},\alpha_{2},\alpha_{3})$
is a 3-vector. For a given $\mbox{\boldmath $\alpha$}$, the two values
$\pm (1 - |\mbox{\boldmath $\alpha$}|^{2})^{\frac{1}{2}}$
of the ``scalar'' $\alpha_{0}$ give the double covering of SO(3).
Then for a given set of signs of the
$\N + 1$ scalars $\alpha_{0}$, the $\N +1$ vectors
$\mbox{\boldmath $\alpha$}$ may be visualized as in fig. 3.

A gauge transformation at the origin of fig. 1 corresponds simply to a
common rotation of all the $\N +1$ vectors
$\mbox{\boldmath $\alpha$}$ in fig. 3, while the
$\N +1$ scalars $\alpha_{0}$ remain unchanged.
Thus, any scalar quantity that
can be formed from the scalars $\alpha_{0}$ and the vectors
$\mbox{\boldmath $\alpha$}$ is gauge-invariant.
But from elementary vector algebra, all
such scalars can be expressed in terms \hfill

\noindent
\begin{picture}(425,290)(0,-50)

\put(200,0){\vector(-1,4){35}}
\put(200,0){\vector( 1,4){40}}
\put(200,0){\vector( 1,2){70}}
\put(200,0){\vector( 3,4){80}}
\put(200,0){\vector( 4,1){100}}

\put(200,0){\vector(1,0){50}}
\put(200,0){\vector(0,1){50}}
\put(200,0){\vector(-1,-1){25}}

\put(160,160){\makebox(0,0){\bf 1}}
\put(245,180){\makebox(0,0){\bf 2}}
\put(280,158){\makebox(0,0){\bf 3}}
\put(292,120){\makebox(0,0){\bf 4}}
\put(322,25){\makebox(0,0){\boldmath $\N$}}
\put(335,25){\makebox(0,0){\bf +1}}

\put(200,123){\makebox(0,0){$\theta_{12}$}}
\put(216,107){\makebox(0,0){$\theta_{13}$}}
\put(200,95){\makebox(0,0){$\theta_{14}$}}
\put(249,123){\makebox(0,0){$\theta_{23}$}}
\put(260,100){\makebox(0,0){$\theta_{24}$}}

% first arc

\put(175.06,100.04){\circle*{1.0}}
\put(175.46,100.34){\circle*{1.0}}
\put(175.86,100.64){\circle*{1.0}}
\put(176.27,100.93){\circle*{1.0}}
\put(176.67,101.23){\circle*{1.0}}
\put(177.08,101.52){\circle*{1.0}}
\put(177.48,101.81){\circle*{1.0}}
\put(177.89,102.09){\circle*{1.0}}
\put(178.30,102.38){\circle*{1.0}}
\put(178.72,102.66){\circle*{1.0}}
\put(179.13,102.94){\circle*{1.0}}
\put(179.54,103.22){\circle*{1.0}}
\put(179.96,103.50){\circle*{1.0}}
\put(180.38,103.78){\circle*{1.0}}
\put(180.80,104.05){\circle*{1.0}}
\put(181.22,104.32){\circle*{1.0}}
\put(181.64,104.59){\circle*{1.0}}
\put(182.06,104.86){\circle*{1.0}}
\put(182.49,105.12){\circle*{1.0}}
\put(182.91,105.38){\circle*{1.0}}
\put(183.34,105.64){\circle*{1.0}}
\put(183.77,105.90){\circle*{1.0}}
\put(184.20,106.16){\circle*{1.0}}
\put(184.63,106.41){\circle*{1.0}}
\put(185.06,106.66){\circle*{1.0}}
\put(185.49,106.91){\circle*{1.0}}
\put(185.93,107.16){\circle*{1.0}}
\put(186.36,107.40){\circle*{1.0}}
\put(186.80,107.65){\circle*{1.0}}
\put(187.24,107.89){\circle*{1.0}}
\put(187.68,108.13){\circle*{1.0}}
\put(188.12,108.36){\circle*{1.0}}
\put(188.56,108.60){\circle*{1.0}}
\put(189.00,108.83){\circle*{1.0}}
\put(189.45,109.06){\circle*{1.0}}
\put(189.89,109.29){\circle*{1.0}}
\put(190.34,109.51){\circle*{1.0}}
\put(190.79,109.74){\circle*{1.0}}
\put(191.23,109.96){\circle*{1.0}}
\put(191.68,110.17){\circle*{1.0}}
\put(192.13,110.39){\circle*{1.0}}
\put(192.59,110.61){\circle*{1.0}}
\put(193.04,110.82){\circle*{1.0}}
\put(193.49,111.03){\circle*{1.0}}
\put(193.95,111.23){\circle*{1.0}}
\put(194.40,111.44){\circle*{1.0}}
\put(194.86,111.64){\circle*{1.0}}
\put(195.32,111.84){\circle*{1.0}}
\put(195.78,112.04){\circle*{1.0}}
\put(196.24,112.24){\circle*{1.0}}
\put(196.70,112.43){\circle*{1.0}}
\put(197.16,112.62){\circle*{1.0}}
\put(197.62,112.81){\circle*{1.0}}
\put(198.09,113.00){\circle*{1.0}}
\put(198.55,113.18){\circle*{1.0}}
\put(199.02,113.36){\circle*{1.0}}
\put(199.49,113.54){\circle*{1.0}}
\put(199.95,113.72){\circle*{1.0}}
\put(200.42,113.89){\circle*{1.0}}
\put(200.89,114.07){\circle*{1.0}}
\put(201.36,114.24){\circle*{1.0}}
\put(201.83,114.41){\circle*{1.0}}
\put(202.30,114.57){\circle*{1.0}}
\put(202.78,114.73){\circle*{1.0}}
\put(203.25,114.90){\circle*{1.0}}
\put(203.72,115.05){\circle*{1.0}}
\put(204.20,115.21){\circle*{1.0}}
\put(204.67,115.36){\circle*{1.0}}
\put(205.15,115.51){\circle*{1.0}}
\put(205.63,115.66){\circle*{1.0}}
\put(206.11,115.81){\circle*{1.0}}
\put(206.58,115.95){\circle*{1.0}}
\put(207.06,116.10){\circle*{1.0}}
\put(207.54,116.24){\circle*{1.0}}
\put(208.03,116.37){\circle*{1.0}}
\put(208.51,116.51){\circle*{1.0}}
\put(208.99,116.64){\circle*{1.0}}
\put(209.47,116.77){\circle*{1.0}}
\put(209.96,116.90){\circle*{1.0}}
\put(210.44,117.02){\circle*{1.0}}
\put(210.92,117.14){\circle*{1.0}}
\put(211.41,117.26){\circle*{1.0}}
\put(211.90,117.38){\circle*{1.0}}
\put(212.38,117.50){\circle*{1.0}}
\put(212.87,117.61){\circle*{1.0}}
\put(213.36,117.72){\circle*{1.0}}
\put(213.85,117.83){\circle*{1.0}}
\put(214.33,117.93){\circle*{1.0}}
\put(214.82,118.04){\circle*{1.0}}
\put(215.31,118.14){\circle*{1.0}}
\put(215.80,118.23){\circle*{1.0}}
\put(216.29,118.33){\circle*{1.0}}
\put(216.79,118.42){\circle*{1.0}}
\put(217.28,118.51){\circle*{1.0}}
\put(217.77,118.60){\circle*{1.0}}
\put(218.26,118.69){\circle*{1.0}}
\put(218.75,118.77){\circle*{1.0}}
\put(219.25,118.85){\circle*{1.0}}
\put(219.74,118.93){\circle*{1.0}}
\put(220.24,119.01){\circle*{1.0}}
\put(220.73,119.08){\circle*{1.0}}
\put(221.23,119.15){\circle*{1.0}}
\put(221.72,119.22){\circle*{1.0}}
\put(222.22,119.29){\circle*{1.0}}
\put(222.71,119.35){\circle*{1.0}}
\put(223.21,119.41){\circle*{1.0}}
\put(223.71,119.47){\circle*{1.0}}
\put(224.20,119.53){\circle*{1.0}}
\put(224.70,119.58){\circle*{1.0}}
\put(225.20,119.63){\circle*{1.0}}
\put(225.69,119.68){\circle*{1.0}}
\put(226.19,119.73){\circle*{1.0}}
\put(226.69,119.77){\circle*{1.0}}
\put(227.19,119.81){\circle*{1.0}}
\put(227.69,119.85){\circle*{1.0}}
\put(228.19,119.89){\circle*{1.0}}
\put(228.68,119.92){\circle*{1.0}}
\put(229.18,119.95){\circle*{1.0}}
\put(229.68,119.98){\circle*{1.0}}

% second arc

\put(175.37,100.23){\circle*{1.0}}
\put(175.80,100.48){\circle*{1.0}}
\put(176.23,100.73){\circle*{1.0}}
\put(176.66,100.98){\circle*{1.0}}
\put(177.10,101.23){\circle*{1.0}}
\put(177.53,101.47){\circle*{1.0}}
\put(177.97,101.72){\circle*{1.0}}
\put(178.41,101.96){\circle*{1.0}}
\put(178.85,102.20){\circle*{1.0}}
\put(179.29,102.43){\circle*{1.0}}
\put(179.73,102.67){\circle*{1.0}}
\put(180.17,102.90){\circle*{1.0}}
\put(180.62,103.13){\circle*{1.0}}
\put(181.06,103.36){\circle*{1.0}}
\put(181.51,103.58){\circle*{1.0}}
\put(181.96,103.81){\circle*{1.0}}
\put(182.40,104.03){\circle*{1.0}}
\put(182.85,104.24){\circle*{1.0}}
\put(183.30,104.46){\circle*{1.0}}
\put(183.76,104.68){\circle*{1.0}}
\put(184.21,104.89){\circle*{1.0}}
\put(184.66,105.10){\circle*{1.0}}
\put(185.12,105.30){\circle*{1.0}}
\put(185.57,105.51){\circle*{1.0}}
\put(186.03,105.71){\circle*{1.0}}
\put(186.49,105.91){\circle*{1.0}}
\put(186.95,106.11){\circle*{1.0}}
\put(187.41,106.31){\circle*{1.0}}
\put(187.87,106.50){\circle*{1.0}}
\put(188.33,106.69){\circle*{1.0}}
\put(188.79,106.88){\circle*{1.0}}
\put(189.26,107.07){\circle*{1.0}}
\put(189.72,107.25){\circle*{1.0}}
\put(190.19,107.43){\circle*{1.0}}
\put(190.66,107.61){\circle*{1.0}}
\put(191.12,107.79){\circle*{1.0}}
\put(191.59,107.96){\circle*{1.0}}
\put(192.06,108.14){\circle*{1.0}}
\put(192.53,108.31){\circle*{1.0}}
\put(193.00,108.48){\circle*{1.0}}
\put(193.47,108.64){\circle*{1.0}}
\put(193.95,108.80){\circle*{1.0}}
\put(194.42,108.97){\circle*{1.0}}
\put(194.89,109.12){\circle*{1.0}}
\put(195.37,109.28){\circle*{1.0}}
\put(195.84,109.43){\circle*{1.0}}
\put(196.32,109.58){\circle*{1.0}}
\put(196.80,109.73){\circle*{1.0}}
\put(197.28,109.88){\circle*{1.0}}
\put(197.75,110.02){\circle*{1.0}}
\put(198.23,110.17){\circle*{1.0}}
\put(198.71,110.31){\circle*{1.0}}
\put(199.20,110.44){\circle*{1.0}}
\put(199.68,110.58){\circle*{1.0}}
\put(200.16,110.71){\circle*{1.0}}
\put(200.64,110.84){\circle*{1.0}}
\put(201.13,110.97){\circle*{1.0}}
\put(201.61,111.09){\circle*{1.0}}
\put(202.09,111.21){\circle*{1.0}}
\put(202.58,111.33){\circle*{1.0}}
\put(203.07,111.45){\circle*{1.0}}
\put(203.55,111.57){\circle*{1.0}}
\put(204.04,111.68){\circle*{1.0}}
\put(204.53,111.79){\circle*{1.0}}
\put(205.02,111.90){\circle*{1.0}}
\put(205.50,112.00){\circle*{1.0}}
\put(205.99,112.11){\circle*{1.0}}
\put(206.48,112.21){\circle*{1.0}}
\put(206.97,112.30){\circle*{1.0}}
\put(207.46,112.40){\circle*{1.0}}
\put(207.96,112.49){\circle*{1.0}}
\put(208.45,112.58){\circle*{1.0}}
\put(208.94,112.67){\circle*{1.0}}
\put(209.43,112.76){\circle*{1.0}}
\put(209.92,112.84){\circle*{1.0}}
\put(210.42,112.92){\circle*{1.0}}
\put(210.91,113.00){\circle*{1.0}}
\put(211.41,113.08){\circle*{1.0}}
\put(211.90,113.15){\circle*{1.0}}
\put(212.40,113.22){\circle*{1.0}}
\put(212.89,113.29){\circle*{1.0}}
\put(213.39,113.36){\circle*{1.0}}
\put(213.88,113.42){\circle*{1.0}}
\put(214.38,113.48){\circle*{1.0}}
\put(214.88,113.54){\circle*{1.0}}
\put(215.37,113.60){\circle*{1.0}}
\put(215.87,113.65){\circle*{1.0}}
\put(216.37,113.70){\circle*{1.0}}
\put(216.86,113.75){\circle*{1.0}}
\put(217.36,113.80){\circle*{1.0}}
\put(217.86,113.84){\circle*{1.0}}
\put(218.36,113.88){\circle*{1.0}}
\put(218.86,113.92){\circle*{1.0}}
\put(219.36,113.96){\circle*{1.0}}
\put(219.85,113.99){\circle*{1.0}}
\put(220.35,114.02){\circle*{1.0}}
\put(220.85,114.05){\circle*{1.0}}
\put(221.35,114.08){\circle*{1.0}}
\put(221.85,114.10){\circle*{1.0}}
\put(222.35,114.12){\circle*{1.0}}
\put(222.85,114.14){\circle*{1.0}}
\put(223.35,114.16){\circle*{1.0}}
\put(223.85,114.17){\circle*{1.0}}
\put(224.35,114.18){\circle*{1.0}}
\put(224.85,114.19){\circle*{1.0}}
\put(225.35,114.20){\circle*{1.0}}
\put(225.85,114.20){\circle*{1.0}}
\put(226.35,114.20){\circle*{1.0}}
\put(226.85,114.20){\circle*{1.0}}
\put(227.35,114.20){\circle*{1.0}}
\put(227.85,114.19){\circle*{1.0}}
\put(228.35,114.18){\circle*{1.0}}
\put(228.85,114.17){\circle*{1.0}}
\put(229.35,114.16){\circle*{1.0}}
\put(229.85,114.14){\circle*{1.0}}
\put(230.35,114.12){\circle*{1.0}}
\put(230.85,114.10){\circle*{1.0}}
\put(231.35,114.08){\circle*{1.0}}
\put(231.85,114.05){\circle*{1.0}}
\put(232.35,114.02){\circle*{1.0}}
\put(232.85,113.99){\circle*{1.0}}
\put(233.34,113.96){\circle*{1.0}}
\put(233.84,113.92){\circle*{1.0}}
\put(234.34,113.88){\circle*{1.0}}
\put(234.84,113.84){\circle*{1.0}}
\put(235.34,113.80){\circle*{1.0}}
\put(235.84,113.75){\circle*{1.0}}
\put(236.33,113.70){\circle*{1.0}}
\put(236.83,113.65){\circle*{1.0}}
\put(237.33,113.60){\circle*{1.0}}
\put(237.82,113.54){\circle*{1.0}}
\put(238.32,113.48){\circle*{1.0}}
\put(238.82,113.42){\circle*{1.0}}
\put(239.31,113.36){\circle*{1.0}}
\put(239.81,113.29){\circle*{1.0}}
\put(240.30,113.22){\circle*{1.0}}
\put(240.80,113.15){\circle*{1.0}}
\put(241.29,113.08){\circle*{1.0}}
\put(241.79,113.00){\circle*{1.0}}
\put(242.28,112.92){\circle*{1.0}}
\put(242.78,112.84){\circle*{1.0}}
\put(243.27,112.76){\circle*{1.0}}
\put(243.76,112.67){\circle*{1.0}}
\put(244.25,112.58){\circle*{1.0}}
\put(244.74,112.49){\circle*{1.0}}
\put(245.24,112.40){\circle*{1.0}}
\put(245.73,112.30){\circle*{1.0}}
\put(246.22,112.21){\circle*{1.0}}
\put(246.71,112.11){\circle*{1.0}}
\put(247.20,112.00){\circle*{1.0}}
\put(247.68,111.90){\circle*{1.0}}
\put(248.17,111.79){\circle*{1.0}}
\put(248.66,111.68){\circle*{1.0}}
\put(249.15,111.57){\circle*{1.0}}
\put(249.63,111.45){\circle*{1.0}}
\put(250.12,111.33){\circle*{1.0}}
\put(250.61,111.21){\circle*{1.0}}
\put(251.09,111.09){\circle*{1.0}}
\put(251.57,110.97){\circle*{1.0}}
\put(252.06,110.84){\circle*{1.0}}
\put(252.54,110.71){\circle*{1.0}}
\put(253.02,110.58){\circle*{1.0}}
\put(253.50,110.44){\circle*{1.0}}
\put(253.99,110.31){\circle*{1.0}}
\put(254.47,110.17){\circle*{1.0}}
\put(254.95,110.02){\circle*{1.0}}

% third arc

\put(175.07,100.00){\circle*{1.0}}
\put(175.56,100.11){\circle*{1.0}}
\put(176.04,100.21){\circle*{1.0}}
\put(176.53,100.32){\circle*{1.0}}
\put(177.02,100.42){\circle*{1.0}}
\put(177.51,100.51){\circle*{1.0}}
\put(178.00,100.61){\circle*{1.0}}
\put(178.50,100.70){\circle*{1.0}}
\put(178.99,100.79){\circle*{1.0}}
\put(179.48,100.88){\circle*{1.0}}
\put(179.97,100.97){\circle*{1.0}}
\put(180.46,101.05){\circle*{1.0}}
\put(180.96,101.13){\circle*{1.0}}
\put(181.45,101.21){\circle*{1.0}}
\put(181.95,101.29){\circle*{1.0}}
\put(182.44,101.36){\circle*{1.0}}
\put(182.94,101.43){\circle*{1.0}}
\put(183.43,101.50){\circle*{1.0}}
\put(183.93,101.57){\circle*{1.0}}
\put(184.42,101.63){\circle*{1.0}}
\put(184.92,101.69){\circle*{1.0}}
\put(185.42,101.75){\circle*{1.0}}
\put(185.91,101.81){\circle*{1.0}}
\put(186.41,101.86){\circle*{1.0}}
\put(186.91,101.91){\circle*{1.0}}
\put(187.40,101.96){\circle*{1.0}}
\put(187.90,102.01){\circle*{1.0}}
\put(188.40,102.05){\circle*{1.0}}
\put(188.90,102.09){\circle*{1.0}}
\put(189.40,102.13){\circle*{1.0}}
\put(189.90,102.17){\circle*{1.0}}
\put(190.39,102.20){\circle*{1.0}}
\put(190.89,102.23){\circle*{1.0}}
\put(191.39,102.26){\circle*{1.0}}
\put(191.89,102.29){\circle*{1.0}}
\put(192.39,102.31){\circle*{1.0}}
\put(192.89,102.33){\circle*{1.0}}
\put(193.39,102.35){\circle*{1.0}}
\put(193.89,102.37){\circle*{1.0}}
\put(194.39,102.38){\circle*{1.0}}
\put(194.89,102.39){\circle*{1.0}}
\put(195.39,102.40){\circle*{1.0}}
\put(195.89,102.41){\circle*{1.0}}
\put(196.39,102.41){\circle*{1.0}}
\put(196.89,102.41){\circle*{1.0}}
\put(197.39,102.41){\circle*{1.0}}
\put(197.89,102.41){\circle*{1.0}}
\put(198.39,102.40){\circle*{1.0}}
\put(198.89,102.39){\circle*{1.0}}
\put(199.39,102.38){\circle*{1.0}}
\put(199.89,102.37){\circle*{1.0}}
\put(200.39,102.35){\circle*{1.0}}
\put(200.89,102.33){\circle*{1.0}}
\put(201.39,102.31){\circle*{1.0}}
\put(201.89,102.29){\circle*{1.0}}
\put(202.39,102.26){\circle*{1.0}}
\put(202.89,102.23){\circle*{1.0}}
\put(203.39,102.20){\circle*{1.0}}
\put(203.88,102.17){\circle*{1.0}}
\put(204.38,102.13){\circle*{1.0}}
\put(204.88,102.09){\circle*{1.0}}
\put(205.38,102.05){\circle*{1.0}}
\put(205.88,102.01){\circle*{1.0}}
\put(206.38,101.96){\circle*{1.0}}
\put(206.87,101.91){\circle*{1.0}}
\put(207.37,101.86){\circle*{1.0}}
\put(207.87,101.81){\circle*{1.0}}
\put(208.36,101.75){\circle*{1.0}}
\put(208.86,101.69){\circle*{1.0}}
\put(209.36,101.63){\circle*{1.0}}
\put(209.85,101.57){\circle*{1.0}}
\put(210.35,101.50){\circle*{1.0}}
\put(210.84,101.43){\circle*{1.0}}
\put(211.34,101.36){\circle*{1.0}}
\put(211.83,101.29){\circle*{1.0}}
\put(212.33,101.21){\circle*{1.0}}
\put(212.82,101.13){\circle*{1.0}}
\put(213.32,101.05){\circle*{1.0}}
\put(213.81,100.97){\circle*{1.0}}
\put(214.30,100.88){\circle*{1.0}}
\put(214.79,100.79){\circle*{1.0}}
\put(215.28,100.70){\circle*{1.0}}
\put(215.78,100.61){\circle*{1.0}}
\put(216.27,100.51){\circle*{1.0}}
\put(216.76,100.42){\circle*{1.0}}
\put(217.25,100.32){\circle*{1.0}}
\put(217.74,100.21){\circle*{1.0}}
\put(218.22,100.11){\circle*{1.0}}
\put(218.71,100.00){\circle*{1.0}}
\put(219.20, 99.89){\circle*{1.0}}
\put(219.69, 99.78){\circle*{1.0}}
\put(220.17, 99.66){\circle*{1.0}}
\put(220.66, 99.54){\circle*{1.0}}
\put(221.15, 99.42){\circle*{1.0}}
\put(221.63, 99.30){\circle*{1.0}}
\put(222.11, 99.18){\circle*{1.0}}
\put(222.60, 99.05){\circle*{1.0}}
\put(223.08, 98.92){\circle*{1.0}}
\put(223.56, 98.79){\circle*{1.0}}
\put(224.04, 98.65){\circle*{1.0}}
\put(224.53, 98.52){\circle*{1.0}}
\put(225.01, 98.38){\circle*{1.0}}
\put(225.49, 98.23){\circle*{1.0}}
\put(225.96, 98.09){\circle*{1.0}}
\put(226.44, 97.94){\circle*{1.0}}
\put(226.92, 97.79){\circle*{1.0}}
\put(227.40, 97.64){\circle*{1.0}}
\put(227.87, 97.49){\circle*{1.0}}
\put(228.35, 97.33){\circle*{1.0}}
\put(228.82, 97.18){\circle*{1.0}}
\put(229.29, 97.01){\circle*{1.0}}
\put(229.77, 96.85){\circle*{1.0}}
\put(230.24, 96.69){\circle*{1.0}}
\put(230.71, 96.52){\circle*{1.0}}
\put(231.18, 96.35){\circle*{1.0}}
\put(231.65, 96.17){\circle*{1.0}}
\put(232.12, 96.00){\circle*{1.0}}
\put(232.58, 95.82){\circle*{1.0}}
\put(233.05, 95.64){\circle*{1.0}}
\put(233.52, 95.46){\circle*{1.0}}
\put(233.98, 95.28){\circle*{1.0}}
\put(234.45, 95.09){\circle*{1.0}}
\put(234.91, 94.90){\circle*{1.0}}
\put(235.37, 94.71){\circle*{1.0}}
\put(235.83, 94.52){\circle*{1.0}}
\put(236.29, 94.32){\circle*{1.0}}
\put(236.75, 94.12){\circle*{1.0}}
\put(237.21, 93.92){\circle*{1.0}}
\put(237.67, 93.72){\circle*{1.0}}
\put(238.12, 93.51){\circle*{1.0}}
\put(238.58, 93.31){\circle*{1.0}}
\put(239.03, 93.10){\circle*{1.0}}
\put(239.48, 92.89){\circle*{1.0}}
\put(239.94, 92.67){\circle*{1.0}}
\put(240.39, 92.45){\circle*{1.0}}
\put(240.84, 92.24){\circle*{1.0}}
\put(241.28, 92.02){\circle*{1.0}}
\put(241.73, 91.79){\circle*{1.0}}
\put(242.18, 91.57){\circle*{1.0}}
\put(242.62, 91.34){\circle*{1.0}}
\put(243.07, 91.11){\circle*{1.0}}
\put(243.51, 90.88){\circle*{1.0}}
\put(243.95, 90.64){\circle*{1.0}}
\put(244.39, 90.41){\circle*{1.0}}
\put(244.83, 90.17){\circle*{1.0}}
\put(245.27, 89.93){\circle*{1.0}}
\put(245.71, 89.68){\circle*{1.0}}
\put(246.14, 89.44){\circle*{1.0}}
\put(246.58, 89.19){\circle*{1.0}}
\put(247.01, 88.94){\circle*{1.0}}
\put(247.44, 88.69){\circle*{1.0}}
\put(247.87, 88.44){\circle*{1.0}}
\put(248.30, 88.18){\circle*{1.0}}
\put(248.73, 87.92){\circle*{1.0}}
\put(249.16, 87.66){\circle*{1.0}}
\put(249.58, 87.40){\circle*{1.0}}
\put(250.01, 87.14){\circle*{1.0}}
\put(250.43, 86.87){\circle*{1.0}}
\put(250.85, 86.60){\circle*{1.0}}
\put(251.27, 86.33){\circle*{1.0}}
\put(251.69, 86.06){\circle*{1.0}}
\put(252.11, 85.78){\circle*{1.0}}
\put(252.53, 85.50){\circle*{1.0}}
\put(252.94, 85.22){\circle*{1.0}}
\put(253.35, 84.94){\circle*{1.0}}
\put(253.77, 84.66){\circle*{1.0}}
\put(254.18, 84.37){\circle*{1.0}}
\put(254.59, 84.09){\circle*{1.0}}
\put(254.99, 83.80){\circle*{1.0}}
\put(255.40, 83.51){\circle*{1.0}}
\put(255.80, 83.21){\circle*{1.0}}
\put(256.21, 82.92){\circle*{1.0}}
\put(256.61, 82.62){\circle*{1.0}}
\put(257.01, 82.32){\circle*{1.0}}
\put(257.41, 82.02){\circle*{1.0}}
\put(257.81, 81.71){\circle*{1.0}}
\put(258.20, 81.41){\circle*{1.0}}
\put(258.60, 81.10){\circle*{1.0}}
\put(258.99, 80.79){\circle*{1.0}}
\put(259.38, 80.48){\circle*{1.0}}
\put(259.77, 80.17){\circle*{1.0}}

% fourth arc

\put(230.44,119.89){\circle*{1.0}}
\put(230.92,119.77){\circle*{1.0}}
\put(231.41,119.64){\circle*{1.0}}
\put(231.89,119.51){\circle*{1.0}}
\put(232.37,119.38){\circle*{1.0}}
\put(232.85,119.24){\circle*{1.0}}
\put(233.34,119.11){\circle*{1.0}}
\put(233.82,118.97){\circle*{1.0}}
\put(234.30,118.82){\circle*{1.0}}
\put(234.77,118.68){\circle*{1.0}}
\put(235.25,118.53){\circle*{1.0}}
\put(235.73,118.38){\circle*{1.0}}
\put(236.21,118.23){\circle*{1.0}}
\put(236.68,118.08){\circle*{1.0}}
\put(237.16,117.92){\circle*{1.0}}
\put(237.63,117.77){\circle*{1.0}}
\put(238.10,117.60){\circle*{1.0}}
\put(238.58,117.44){\circle*{1.0}}
\put(239.05,117.28){\circle*{1.0}}
\put(239.52,117.11){\circle*{1.0}}
\put(239.99,116.94){\circle*{1.0}}
\put(240.46,116.76){\circle*{1.0}}
\put(240.93,116.59){\circle*{1.0}}
\put(241.39,116.41){\circle*{1.0}}
\put(241.86,116.23){\circle*{1.0}}
\put(242.33,116.05){\circle*{1.0}}
\put(242.79,115.87){\circle*{1.0}}
\put(243.26,115.68){\circle*{1.0}}
\put(243.72,115.49){\circle*{1.0}}
\put(244.18,115.30){\circle*{1.0}}
\put(244.64,115.11){\circle*{1.0}}
\put(245.10,114.91){\circle*{1.0}}
\put(245.56,114.71){\circle*{1.0}}
\put(246.02,114.51){\circle*{1.0}}
\put(246.48,114.31){\circle*{1.0}}
\put(246.93,114.10){\circle*{1.0}}
\put(247.39,113.90){\circle*{1.0}}
\put(247.84,113.69){\circle*{1.0}}
\put(248.29,113.48){\circle*{1.0}}
\put(248.75,113.26){\circle*{1.0}}
\put(249.20,113.04){\circle*{1.0}}
\put(249.65,112.83){\circle*{1.0}}
\put(250.09,112.61){\circle*{1.0}}
\put(250.54,112.38){\circle*{1.0}}
\put(250.99,112.16){\circle*{1.0}}
\put(251.43,111.93){\circle*{1.0}}
\put(251.88,111.70){\circle*{1.0}}
\put(252.32,111.47){\circle*{1.0}}
\put(252.76,111.23){\circle*{1.0}}
\put(253.20,111.00){\circle*{1.0}}
\put(253.64,110.76){\circle*{1.0}}
\put(254.08,110.52){\circle*{1.0}}
\put(254.52,110.27){\circle*{1.0}}
\put(254.95,110.03){\circle*{1.0}}

% fifth arc

\put(230.03,119.98){\circle*{1.0}}
\put(230.42,119.67){\circle*{1.0}}
\put(230.81,119.35){\circle*{1.0}}
\put(231.19,119.03){\circle*{1.0}}
\put(231.58,118.72){\circle*{1.0}}
\put(231.96,118.39){\circle*{1.0}}
\put(232.34,118.07){\circle*{1.0}}
\put(232.72,117.75){\circle*{1.0}}
\put(233.10,117.42){\circle*{1.0}}
\put(233.48,117.09){\circle*{1.0}}
\put(233.85,116.76){\circle*{1.0}}
\put(234.23,116.43){\circle*{1.0}}
\put(234.60,116.09){\circle*{1.0}}
\put(234.97,115.76){\circle*{1.0}}
\put(235.34,115.42){\circle*{1.0}}
\put(235.70,115.08){\circle*{1.0}}
\put(236.07,114.74){\circle*{1.0}}
\put(236.43,114.39){\circle*{1.0}}
\put(236.79,114.05){\circle*{1.0}}
\put(237.15,113.70){\circle*{1.0}}
\put(237.51,113.35){\circle*{1.0}}
\put(237.87,113.00){\circle*{1.0}}
\put(238.22,112.65){\circle*{1.0}}
\put(238.58,112.29){\circle*{1.0}}
\put(238.93,111.94){\circle*{1.0}}
\put(239.28,111.58){\circle*{1.0}}
\put(239.62,111.22){\circle*{1.0}}
\put(239.97,110.86){\circle*{1.0}}
\put(240.31,110.50){\circle*{1.0}}
\put(240.65,110.13){\circle*{1.0}}
\put(240.99,109.77){\circle*{1.0}}
\put(241.33,109.40){\circle*{1.0}}
\put(241.67,109.03){\circle*{1.0}}
\put(242.00,108.66){\circle*{1.0}}
\put(242.34,108.28){\circle*{1.0}}
\put(242.67,107.91){\circle*{1.0}}
\put(243.00,107.53){\circle*{1.0}}
\put(243.32,107.15){\circle*{1.0}}
\put(243.65,106.77){\circle*{1.0}}
\put(243.97,106.39){\circle*{1.0}}
\put(244.29,106.01){\circle*{1.0}}
\put(244.61,105.63){\circle*{1.0}}
\put(244.93,105.24){\circle*{1.0}}
\put(245.25,104.85){\circle*{1.0}}
\put(245.56,104.46){\circle*{1.0}}
\put(245.87,104.07){\circle*{1.0}}
\put(246.18,103.68){\circle*{1.0}}
\put(246.49,103.28){\circle*{1.0}}
\put(246.80,102.89){\circle*{1.0}}
\put(247.10,102.49){\circle*{1.0}}
\put(247.40,102.09){\circle*{1.0}}
\put(247.70,101.69){\circle*{1.0}}
\put(248.00,101.29){\circle*{1.0}}
\put(248.30,100.89){\circle*{1.0}}
\put(248.59,100.48){\circle*{1.0}}
\put(248.88,100.08){\circle*{1.0}}
\put(249.17, 99.67){\circle*{1.0}}
\put(249.46, 99.26){\circle*{1.0}}
\put(249.74, 98.85){\circle*{1.0}}
\put(250.03, 98.44){\circle*{1.0}}
\put(250.31, 98.03){\circle*{1.0}}
\put(250.59, 97.61){\circle*{1.0}}
\put(250.87, 97.20){\circle*{1.0}}
\put(251.14, 96.78){\circle*{1.0}}
\put(251.42, 96.36){\circle*{1.0}}
\put(251.69, 95.94){\circle*{1.0}}
\put(251.96, 95.52){\circle*{1.0}}
\put(252.22, 95.10){\circle*{1.0}}
\put(252.49, 94.67){\circle*{1.0}}
\put(252.75, 94.25){\circle*{1.0}}
\put(253.01, 93.82){\circle*{1.0}}
\put(253.27, 93.39){\circle*{1.0}}
\put(253.53, 92.96){\circle*{1.0}}
\put(253.78, 92.53){\circle*{1.0}}
\put(254.03, 92.10){\circle*{1.0}}
\put(254.28, 91.67){\circle*{1.0}}
\put(254.53, 91.23){\circle*{1.0}}
\put(254.78, 90.80){\circle*{1.0}}
\put(255.02, 90.36){\circle*{1.0}}
\put(255.26, 89.92){\circle*{1.0}}
\put(255.50, 89.48){\circle*{1.0}}
\put(255.74, 89.04){\circle*{1.0}}
\put(255.97, 88.60){\circle*{1.0}}
\put(256.20, 88.16){\circle*{1.0}}
\put(256.43, 87.71){\circle*{1.0}}
\put(256.66, 87.27){\circle*{1.0}}
\put(256.89, 86.82){\circle*{1.0}}
\put(257.11, 86.38){\circle*{1.0}}
\put(257.33, 85.93){\circle*{1.0}}
\put(257.55, 85.48){\circle*{1.0}}
\put(257.77, 85.03){\circle*{1.0}}
\put(257.98, 84.58){\circle*{1.0}}
\put(258.19, 84.12){\circle*{1.0}}
\put(258.40, 83.67){\circle*{1.0}}
\put(258.61, 83.21){\circle*{1.0}}
\put(258.82, 82.76){\circle*{1.0}}
\put(259.02, 82.30){\circle*{1.0}}
\put(259.22, 81.84){\circle*{1.0}}
\put(259.42, 81.39){\circle*{1.0}}
\put(259.62, 80.93){\circle*{1.0}}
\put(259.81, 80.46){\circle*{1.0}}

% big dots

\put(240.90, 43.90){\circle*{2.0}}
\put(245.08, 39.60){\circle*{2.0}}
\put(248.80, 34.90){\circle*{2.0}}
\put(252.05, 29.85){\circle*{2.0}}
\put(254.77, 24.51){\circle*{2.0}}
\put(256.94, 18.92){\circle*{2.0}}

\put(0,-50){\makebox(0,0)[l]{\footnotesize{
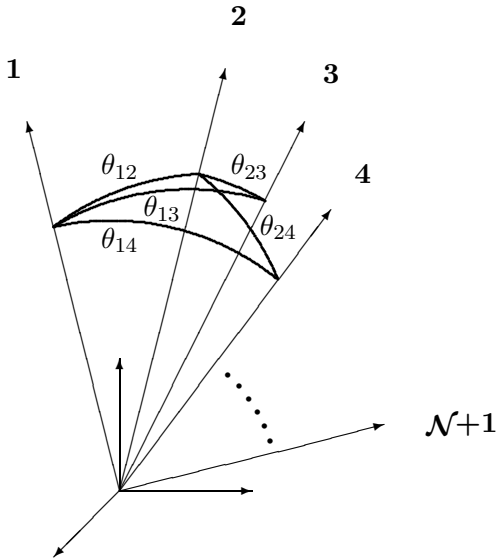
Fig. 3. The $\N + 1$ vectors $\alpha$.
}}}

\end{picture}

\noindent
of the $\alpha_{0}$ and the scalar products and scalar
triple products among the vectors. This is the hint
for the choice of a more appropriate set of \gi variables:
instead of working directly with the traces formed
from these basis matrices, we consider the
so-called $L$ variables \cite{loll} defined here as
\bea
L(\alpha)
&=&
\alpha_{0} \\
&=&
+{\textstyle\frac{1}{2}}{\rm Tr}\,U_{\alpha} \\
L(\alpha,\beta)
&=&
\mbox{\boldmath $\alpha\cdot\beta$} \\
&=&
-{\textstyle\frac{1}{2}}{\rm Tr}\,U_{\alpha}U_{\beta}
+{\textstyle\frac{1}{4}}{\rm Tr}\,U_{\alpha}{\rm Tr}\,U_{\beta} \\
L(\alpha,\beta,\gamma)
&=&
\mbox{\boldmath $\alpha\times\beta\cdot\gamma$} \\
&=&
-{\textstyle\frac{1}{2}}{\rm Tr}\,U_{\alpha}U_{\beta}U_{\gamma}
+{\textstyle\frac{1}{4}}{\rm Tr}\,U_{\alpha}{\rm Tr}\,U_{\beta}U_{\gamma}
+{\textstyle\frac{1}{4}}{\rm Tr}\,U_{\beta}{\rm Tr}\,U_{\gamma}U_{\alpha}
\nonumber \\
& &
+{\textstyle\frac{1}{4}}{\rm Tr}\,U_{\gamma}{\rm Tr}\,U_{\alpha}U_{\beta}
-{\textstyle\frac{1}{4}}{\rm Tr}\,U_{\alpha}{\rm Tr}\,U_{\beta}
{\rm Tr}\,U_{\gamma}
\eea
In terms of these variables, the first three of the six types of
trace identity alluded to above become trivial:
$L(\alpha) = L(\alpha^{-1})$,
$L(\alpha,\beta) = -L(\alpha,\beta^{-1})$ and
$L(\alpha,\beta,\gamma) = -L(\beta,\alpha,\gamma)$.

The fact that it is possible to express any scalar
and hence \gi quantity formed from the scalars $\alpha_{0}$ and vectors
$\mbox{\boldmath $\alpha$}$ in terms
of the $L$ variables is expressed in the fourth type of identity:
\bea
{\rm Tr}\,U_{\alpha}U_{\beta}U_{\gamma}U_{\gamma}
&=&
\tfrac{1}{2}\Bigl(
  L(\alpha) L(\beta,\gamma,\delta)
+ L(\beta)  L(\gamma,\delta,\alpha)
+ L(\gamma) L(\delta,\alpha,\beta)
+ L(\delta) L(\alpha,\beta,\gamma) \nonumber \\
& &
+ L(\alpha,\beta)L(\gamma,\delta)
+ L(\alpha,\gamma)L(\beta,\delta)
+ L(\alpha,\delta)L(\beta,\gamma)
\Bigr) \nonumber \\
& &
+\tfrac{1}{4}\Bigl(
  L(\alpha)L(\beta)L(\gamma,\delta)
+ L(\beta)L(\gamma)L(\delta,\alpha)
+ L(\gamma)L(\delta)L(\alpha,\beta) \nonumber \\
& &
+ L(\delta)L(\alpha)L(\beta,\gamma)
+ L(\alpha)L(\gamma)L(\beta,\delta)
+ L(\beta)L(\delta)L(\alpha,\gamma)
\Bigr) \nonumber \\
\label{3.7}
& &
+\tfrac{1}{8}\Bigl(
L(\alpha)L(\beta)L(\gamma)L(\delta)
\Bigr)
\eea
Using the definitions of the $L$ variables,
this identity enables the trace of any product of four or more basis
matrices and their inverses to be written directly in terms of $L$
variables. It amounts simply to multiplying out such a trace in terms
of the scalar and vector components $\alpha_{0}$,
$\mbox{\boldmath $\alpha$}$ of the matrices and using
elementary identities from vector algebra to write all terms in terms
of the $L$ variables.

However, there are still
many more $L$ variables than the $3\N$ physical degrees of
freedom of the system,
so that it is necessary to work out which of them are independent.
The two crucial non-trivial identities required to do this were first
derived by Loll \cite{loll}.

Firstly,
\bea
L(\alpha,\beta,\gamma)^{2}
+L(\alpha,\beta)^{2}L(\gamma,\gamma)
+L(\beta,\gamma)^{2}L(\alpha,\alpha)
+L(\gamma,\alpha)^{2}L(\beta,\beta)
& & \nonumber \\
-2L(\alpha,\beta)L(\beta,\gamma)L(\gamma,\alpha)
-L(\alpha,\alpha)L(\beta,\beta)L(\gamma,\gamma)
&=&
0
\eea
This is the fifth type of identity, whose solution may be written
\be
L(\alpha,\beta,\gamma)
=
s(\alpha,\beta,\gamma)
|L(\alpha,\beta,\gamma)|
\ee
The discrete variable $s(\alpha,\beta,\gamma)$ is
given simply by the sign of $L(\alpha,\beta,\gamma)$
and so has value $+1$ ($-1$) if the angle
between $\mbox{\boldmath $\alpha\times\beta$}$ and
$\mbox{\boldmath $\gamma$}$ is less than (greater than) $90^{\rm o}$
(for $\mbox{\boldmath $\alpha\times\beta\cdot\gamma$} = 0$
it is undefined). The modulus $|L(\alpha,\beta,\gamma)|$
is fully given from eq. (11) by the variables $L(\alpha,\beta)$.
This identity expresses the fact that, given the angles
$\theta_{\alpha\beta}$, $\theta_{\beta\gamma}$, $\theta_{\gamma\alpha}$
between three vectors $\mbox{\boldmath $\alpha$}$,
$\mbox{\boldmath $\beta$}$, $\mbox{\boldmath $\gamma$}$, the relative
orientation of the three vectors is specified up to a possible
``reflection'' of, say, $\mbox{\boldmath $\gamma$}$
in the plane formed by $\mbox{\boldmath $\alpha$}$ and
$\mbox{\boldmath $\beta$}$. This information is given by the sign
of $L(\alpha,\beta,\gamma)$. Thus, the $L(\alpha,\beta,\gamma)$ can be
be fully expressed in terms of the $L(\alpha,\beta)$ and the
discrete variables $s(\alpha,\beta,\gamma)$. From fig. 3, there
are $\N -1$ independent $s(\alpha,\beta,\gamma)$.

Secondly \cite{loll},
\bea
 L(\alpha,\beta)^{2}L(\gamma,\delta)^{2}
+L(\alpha,\gamma)^{2}L(\beta,\delta)^{2}
+L(\alpha,\delta)^{2}L(\beta,\gamma)^{2}
& & \\ \nonumber
-L(\alpha,\beta)^{2}L(\gamma,\gamma)L(\delta,\delta)
-L(\alpha,\gamma)^{2}L(\beta,\beta)L(\delta,\delta)
-L(\alpha,\delta)^{2}L(\beta,\beta)L(\gamma,\gamma)
& & \\ \nonumber
-L(\beta,\gamma)^{2}L(\alpha,\alpha)L(\delta,\delta)
-L(\beta,\delta)^{2}L(\alpha,\alpha)L(\gamma,\gamma)
-L(\gamma,\delta)^{2}L(\alpha,\alpha)L(\beta,\beta)
& & \\ \nonumber
+2L(\alpha,\alpha)L(\beta,\gamma)L(\gamma,\delta)L(\delta,\beta)
+2L(\beta,\beta)L(\alpha,\gamma)L(\gamma,\delta)L(\delta,\alpha)
& & \\ \nonumber
+2L(\gamma,\gamma)L(\alpha,\beta)L(\beta,\delta)L(\delta,\alpha)
+2L(\delta,\delta)L(\alpha,\beta)L(\beta,\gamma)L(\gamma,\alpha)
& & \\ \nonumber
-2L(\alpha,\beta)L(\beta,\gamma)L(\gamma,\delta)L(\delta,\alpha)
-2L(\alpha,\gamma)L(\beta,\gamma)L(\beta,\delta)L(\delta,\alpha)
& & \\ \nonumber
-2L(\alpha,\beta)L(\beta,\delta)L(\gamma,\delta)L(\gamma,\alpha)
+L(\alpha,\alpha)L(\beta,\beta)L(\gamma,\gamma)L(\delta,\delta)
&=&
0
\eea
This is the sixth type of identity, and involves just the variables
$L(\alpha,\beta)$. This identity expresses the fact that, given
four vectors
$\mbox{\boldmath $\alpha$}$,
$\mbox{\boldmath $\beta$}$,
$\mbox{\boldmath $\gamma$}$,
$\mbox{\boldmath $\delta$}$,
one of the six angles between them  may
always be expressed in terms of the other five
and the appropriate discrete variables $s(\alpha,\beta,\gamma)$.

How many independent $L(\alpha)$ and $L(\alpha,\beta)$ do these
identities leave?
Clearly, the $\N +1$ variables $L(\alpha)$ are independent.
The lengths of the vectors are then given via
$|\mbox{\boldmath $\alpha$}| = +(1 - L(\alpha)^{2})^{\frac{1}{2}}$.
Given the vectors' lengths, the angles $\theta_{\alpha\beta}$
between them are given by the $L(\alpha,\beta)$.
{}From fig. 3, there are
$2\N -1$ independent such angles. Thus,
\bea\label{3.8}
{\rm no.\,\,of\,\,independent\,\,} L(\alpha)
&=&
\N + 1 \\
\label{3.9}
{\rm no.\,\,of\,\,independent\,\,} L(\alpha,\beta)
&=&
2\N -1
\eea
giving exactly the correct total number $3\N$ of continuous variables.

Fuller details of these identities and their
solution are given in \cite{watson}.

\pagebreak

{\bf 4. Summary and Conclusions}

\noindent
Thus, by constructing a set of basis matrices $U_{\alpha}$,
$\alpha = 1,2 \ldots \N+1$,
for a given SU(2) lattice gauge theory and expressing
them in the form eq. (2),
the fact that the \gi quantities are the scalar quantities
which can be formed from the scalars $\alpha_{0}$ and the vectors
$\mbox{\boldmath $\alpha$}$, and that all such scalars can be
expressed in terms of the $\alpha_{0}$ and the scalar products and
scalar triple products among the vectors, motivates the change of
variables from the Wilson and Polyakov loops themselves to the $L$
variables eqs. (4)-(9). In terms of these $L$ variables, the content
of the SU(2) Mandelstam constraints becomes transparent
-- they are just a set of identities in elementary vector algebra --
and their
solution is conceptually straightforward (although it remains
algebraically complicated). Furthermore, it becomes clear that a set
of $3\N$ independent continuous variables
$L(\alpha)$, $L(\alpha,\beta)$, and hence the $3\N$
Wilson and Polyakov
loops themselves from which they are formed, are not
enough, {\em even in some local region}, to describe the subspace of
physical configurations, but that a set of $\N -1$ \gi discrete
$\pm 1$ variables $s(\alpha,\beta,\gamma)$ are also required.
These discrete variables are
completely unrelated either to the global $Z_{2}$ information
carried in the Polyakov loops (in the example in fig. 1, the variables
${\rm Tr}\,U_{1} \equiv 2L(1)$, ${\rm Tr}\,U_{2} \equiv 2L(2)$) or the
double covering of SO(3) by SU(2) (in general, the two possible signs
of each of the $\alpha_{0} \equiv L(\alpha)$ for a given set of
\nolinebreak vectors \nolinebreak $\mbox{\boldmath $\alpha$}$).

Although the variables constructed after solving the SU(2) Mandelstam
constraints are \gi and involve the correct number $3\N$ of continuous
degrees of freedom, they are unfortunately
neither (quasi-)local nor translation-
and $90^{\rm o}$ rotation-invariant. This is a severe drawback, as it
hinders writing the Hamiltonian in terms of them, and also
makes intractable the jacobian required for the measure.
Further progress almost certainly depends on restoring somehow
the variables' spatial symmetries.
However, in contrast to the more indirect approach taken in
[5-7], by actually
solving the SU(2) Mandelstam constraints direct insight is obtained
into the information carried by the variables and
in particular the role of and
need for the discrete variables $s(\alpha,\beta,\gamma)$.

\vspace*{10pt}

I thank R. Loll for useful discussions.


\begin{thebibliography}{99}

\bibitem{watson} N.J. Watson, Phys. Lett. B323 (1994) 385.

\bibitem{mullerruhl} V. M\"uller and W. Ruhl, Nucl. Phys. B230 (1984) 49.

\bibitem{mandelstam1} S. Mandelstam, Phys. Rev. 175 (1968) 1580.

\bibitem{mandelstam2} S. Mandelstam, Phys. Rev. D19 (1979) 2391.

\bibitem{loll} R. Loll, Nucl. Phys. B368 (1992) 121.

\bibitem{loll2} R. Loll, Nucl. Phys. B (Proc. Sup.) 30 (1993) 224.

\bibitem{loll3} R. Loll, Nucl. Phys. B400 (1993) 126.

\end{thebibliography}
\end{document}